\pdfoutput=1

\documentclass[twocolumn, times]{aastex63} 

\usepackage{natbib}
\usepackage{amsmath}
\usepackage{amssymb}
\usepackage{subfigure}
\usepackage{url}
\usepackage{longtable}

\renewcommand{\vec}[1]{ {\mathbf #1} }

\newcommand{\Eqs}{{Equations}}
\newcommand{\Fig}{{Figure}}

\newcommand{\SDO}{{\it SDO}}

\graphicspath{figs/}

\submitjournal{\textit{ApJ}}
\shorttitle{Pre-flare magnetic energy and helicity}
\shortauthors{Duan et al.}

\begin{document}

\title{A Study of Pre-flare Solar Coronal Magnetic Fields: Magnetic Energy and Helicity}

\author{Aiying Duan} \affiliation{Planetary
  Environmental and Astrobiological Research Laboratory (PEARL),
  School of Atmospheric Sciences, Sun Yat-sen University, Zhuhai
  519000, China; \textcolor[rgb]{0.00,0.07,1.00}{duanaiy@mail.sysu.edu.cn}}
  \affiliation{Key Laboratory of Tropical Atmosphere-Ocean System, Sun Yat-sen University, Ministry of Education, Zhuhai 519000, China}

\author{Chaowei Jiang} \affiliation{Institute of Space Science and
  Applied Technology, Harbin Institute of Technology, Shenzhen 518055,
  China}

\author{Xueshang Feng} \affiliation{Institute of Space Science and Applied Technology, Harbin Institute of Technology, Shenzhen 518055,
  China}

\begin{abstract}
  Solar flares fall into two types with eruptive ones associated with coronal mass ejection (CME) and confined ones without CME. To explore whether there are pre-flare conditions in terms of magnetic energy and helicity that can effectively determine the types of flares, here we analyzed a suite of related parameters of the reconstructed pre-flare coronal magnetic field of major solar flares, either eruptive or confined, from 2011 to 2017 near the solar disk center. The investigated parameters include the extensive-type quantities such as the total magnetic energy $E_T$, the potential energy $E_P$, the free energy $E_F$, the relative helicity $H_R$, and the non-potential helicity $H_J$, as well as the intensive-type indices $E_F/E_P$, $|H_J/H_R|$, $|H_R/\phi^{\prime2}|$ and $|H_J/\phi^{\prime2}|$, where $\phi^{\prime}$ is half of the total unsigned magnetic flux. We have the following key findings: (1) None of the extensive parameters can effectively distinguish the eruptive and confined potential of the pre-flare coronal fields, though the confined events have averagely larger values; (2) All the intensive parameters have significantly larger average and median values for eruptive flares than the confined events, which indicates that the field for eruptive flares have overall higher degree of non-potentiality and complexity than that of the confined flares; (3) The energy ratio $E_F/E_P$ and the normalized non-potential helicity $|H_J/\phi^{\prime2}|$, which are strongly correlated with each other, have among the highest capability of distinguishing the fields that possibly produce a major eruptive or confined flare, as over 75\% of all the events are successfully discriminated between eruptive and confined flares by using critical values of $E_F/E_P\ge0.27$ and $|H_J/\phi^{\prime2}|\ge0.009$.
\end{abstract}


\section{Introduction}
\label{sec:intro}
Solar eruptions, mainly including solar flares and coronal mass ejections (CMEs), are violent activity phenomena occurring in the solar atmosphere. During flares, a sudden and catastrophic energy release occurs in a localized region within tens of minutes, producing an almost entire electromagnetic spectrum enhancement\textbf, as well as plasma heating and particle acceleration~\citep{Fletcher2011, Benz2017}. On the other hand, CMEs carry massive plasma and embedded magnetic field into the interplanetary space, thus an Earth-directed CME can interact with the geomagnetic field and cause hazardous space weather. It is now well established that flares and CMEs are different manifestations of explosive release of magnetic free energy in the corona, and they are often associated with each other, but not one-to-one correspondences. Studies show that most of the major flares are accompanied with CMEs (i.e., eruptive flares), while a large number of flares does not (i.e., confined flares). Until now, although many works~\citep[e.g.,][]{Andrews2003, Yashiro2006, WangYM2007, Toriumi2017, Baumgartner2018, JingJ2018, DuanA2019, LiT2020} have been done, the key factor that causes the difference between the two types of flares is still undetermined.

Since solar eruptions have their root in the evolution of magnetic field in solar atmosphere, many parameters qualifying the magnetic complexity and non-potentiality are investigated, which mainly include the twist~\citep{Pevtsov1994, Hagino2004}, electric current~\citep{Leka1993, WangHM1994SP},  magnetic shear~\citep{WangHM1994ApJ}, horizontal gradient of longitudinal magnetic field~\citep{TianLR2002}, free magnetic energy~\citep{Metcalf2005, JingJ2010, Gupta2021, LiT2021} and magnetic helicity~\citep{Zuccarello2011, Zuccarello2018, Vasantharaju2018, Thalmann2019A, Thalmann2019B, Gupta2021, LiuY2023}, etc. Essentially all the parameters are more or less related to the two of them, namely the free magnetic energy which is the only energy that can power the eruption, and magnetic helicity which is a global measure of the complexity of the specific field configurations. Thus, these two more intrinsic physical parameters got more attentions in the relevant studies.

The free magnetic energy $E_F$ measures the energy deviation of the coronal magnetic field from its potential (i.e., current-free) state $E_P$, and it is regarded as the upper limit of the energy that is available in a coronal magnetic configuration for conversion into kinetic and thermal energies~\citep{Jingj2009}. Therefore, studying of $E_F$ and its evolution plays an important role in understanding the energy storage and release process during solar eruptions. To obtain $E_F$, it is often required to extrapolate the coronal magnetic field from the observed photospheric vector magnetograms. For example, \citet{JingJ2010} showed that there was a positive correlation between the $E_F$ and the flare index. Based on 38 solar eruptive events, \citet{Emslie2012} found that the $E_F$ was sufficient to power the flare-accelerated particles, the hot thermal plasma, and the CME. \citet{Vasantharaju2018} demonstrated for a sample of 77 flare/CME cases that the amount of $E_F$ and its temporal variation were highly correlated to the flare strength and CME speed. However, as an extensive quantity, the $E_F$ has its limitation in flare prediction. For instance, \citet{Gilchrist2012} suggested that the certain content of the pre-flare free energy is not decisive to the occurring of a flare since it is not uniquely related to the complexity of the flare-involved magnetic field.

Magnetic helicity, which measures the overall degree of the geometrical complexity of the magnetic fields including the twist of the field lines as well as the mutual linkages of different field lines, is another parameter relevant to  the non-potentiality of the coronal field~\citep{Pariat2017, Thalmann2019A}. The magnetic helicity of a field $\vec B$ within a volume $V$ is defined as $H_M=\int_V \vec A \cdot \vec B dV$, where $\vec A$ is the vector potential and satisfies $\vec B=\nabla \times \vec A$. This definition of helicity is physically meaningful only for magnetically closed systems (that is, without field lines passing through the surface of $V$). For applications to the solar corona (which is an open system), the relative magnetic helicity $H_R$ was defined by~\citet{Berger1984} and ~\citet{Finn1984} as a gauge invariant form of magnetic helicity with respect to a reference magnetic field, which is naturally given by the potential field $E_P$. In addition, the relative helicity $H_R$ can be further decomposed to the non-potential (or current-carrying) helicity $H_J$ and the volume-threading helicity $H_{PJ}$~\citep{Berger1999, Linan2018}. Several works suggested that the eruptivity of an active region (AR) was closely related to its $H_R$~\citep[e.g.,][]{Nindos2004, LaBonte2007, Smyrli2010, Tziotziou2012} and/or $H_J$~\citep{Pariat2017} content, since eruptive flares seem containing higher values of pre-flare $H_R$ and/or $H_J$ and the total $H_R$ of the AR decreased after eruption~\citep{Bleybel2002}.

However, by analyzing the pre-flare magnetic conditions of three different ARs which produced different types of flares, ~\citet{SunXD2015} concluded that compared to the extensive-type parameters (such as sunspot area, total magnetic flux, electric current, magnetic energy, magnetic helicity, etc.), the intensive-type parameters regarding non-potentiality of the core field (such as the ratio of free energy $E_F$ to potential energy $E_P$) were more important to discriminate the eruptive and the confined flares. Recently, based on both MHD simulations~\citep{Pariat2017, Zuccarello2018} and observations~\citep{James2018, Moraitis2019B, Thalmann2019B, Thalmann2020, Gupta2021}, a number of works were devoted to determining the thresholds of the intensive-type parameters which can distinguish different types of flares. In particular, a new intensive-type parameter $|H_J/H_R|$ is found to show a strong capability to indicate the eruptive potential of an AR. For example, simulations found that the values of $|H_J/H_R|$ for eruptive cases were larger than 0.45~\citep{Pariat2017} or 0.3~\citep{Zuccarello2018}, while observational studies showed much smaller values from 0.1 to 0.17~\citep{James2018, Thalmann2019B}. Furthermore, \citet{Gupta2021} also demonstrated that an AR with $\langle|H_J/H_R|\rangle>0.1$, $\langle E_F/E_T\rangle > 0.2$ ($E_T$ is the total magnetic energy) and $\langle|H_J/\phi^{\prime2}|\rangle > 0.005$ ($\phi^{\prime}$ is the half of the total unsigned magnetic flux) is likely to produce large eruptive flares.

The aforementioned conclusions are mostly based on single-case analysis or small-sample studies, and require to be tested with investigations of larger samples. Moreover, the key parameters are calculated from coronal magnetic field extrapolations, and thus the inferred results might strongly depend on the quality or reliability of the coronal magnetic field extrapolations. While most of the previous studies used the nonlinear force-free field (NLFFF) extrapolation code based on the optimization method as originally developed by~\citet{Wiegelmann2004}, there are currently many other methods available for NLFFF extrapolations from the vector magnetograms. Since it has been shown that different methods seem to produce rather inconsistent results between each other~\citep[e.g.,][]{DeRosa2009, Regnier2013, Aschwanden2014, DuanA2017, Wiegelmann2017}, any results based on any single NLFFF code must be taken with caution, and more studies with different independent codes are valuable and also necessary for a better inspection. In this paper, we performed a statistical study to better understand the roles of different quantities in characterizing the eruptive or confined potential of ARs. We employed the coronal magnetic field extrapolation code developed by~\citet{JiangC2013NLFFF}, named the CESE--MHD--NLFFF code which is based on the MHD-relaxation approach, to study a larger sample of 45 major flares with 29 eruptive and 16 confined. Our results support that the extensive-type parameters cannot effectively distinguish the eruptive and confined potential of the pre-flare coronal fields, though the confined events have averagely larger values. The intensive-type parameters have significantly larger average and median values for eruptive flares than the confined events, which indicates that the field for eruptive flares have overall higher degree of non-potentiality and complexity than that of the confined flares. Comparing to other intensive-type parameters, we found that both $E_F/E_P$ and $|H_J/\phi^{\prime2}|$  play much more important roles in discriminating the eruptive and confined flares. The rest of the paper is organized as follows: The data and method are presented in Section~\ref{sec:data}, the results are given in Section~\ref{sec:res}, and finally a summary is provided in Section~\ref{sec:con}.

\section{Data and Method}
\label{sec:data}
In our previous study~\citep{DuanA2019}, we have performed a statistical survey of the pre-flare specific magnetic configurations for a sample of major solar flares of either eruptive or confined, with focus on the topology and ideal instability of magnetic flux rope. In this study, the same sample is used, which has in total 45 flares from 30 different ARs, including 29 eruptive ones (above GOES-class M5) and 16 confined ones (above M3.9). The flares are observed by the Solar Dynamic Observatory~\citep[{\SDO};][]{Pesnell2012} from 2011 January to 2017 December, and all of them occurred within $45^{\circ}$ in longitude of the disk center (details of the events are listed in Table~\ref{tab:event_list}). For more information about the criterion for events selection, one can refer to~\citet{DuanA2019} and~\citet{DuanA2021apjl}.

\begin{table*}[htbp]
\footnotesize
  \centering
  \caption{List of events and their pre-flare parameters.}
  \begin{tabular}{ccccccccccc}
    \hline
    \hline
    No. & Flare peak time & Flare class  & NOAA AR &  E/C$^{a}$ & $\phi^{\prime}$  & $E_T$  & $E_{F1}$  & $E_{F2}$  & $H_R$  & $H_J$\\
           \ & \ & \ & \ & \ & ($10^{22}$Mx) & ($10^{33}$erg) & ($10^{32}$erg)  & ($10^{32}$erg) & ($10^{43}$Mx$^2$) & ($10^{42}$Mx$^2$)\\
    \hline
    1   & SOL2011-02-13T17:38   &   M6.6   & 11158 &  E & 1.03 & 0.56   & 1.43   & 1.51   & 0.06     & 1.12\\
    2   & SOL2011-02-15T01:56   &   X2.2   &  11158 &  E & 1.43 & 1.09   & 3.04   & 3.19   & 0.77    & 2.88 \\
    3   & SOL2011-03-09T23:23   &   X1.5   &  11166 &  C & 1.89 & 1.54   & 2.06   & 2.21   & 0.48    & 0.44\\
    4   & SOL2011-07-30T02:09   &  M9.3   &  11261 &  C & 1.75 & 0.75   & 1.24   & 1.32   & -0.05    & -0.49\\
    5   & SOL2011-08-03T13:48   &  M6.0   &  11261 &  E & 1.67 & 0.67   & 2.28   & 2.40   & 1.04     & 3.22\\
    6   & SOL2011-09-06T01:50  &  M5.3   &   11283 &  E & 1.40 & 0.66   & 1.17   & 1.24   & 0.26     & 0.39\\
    7   & SOL2011-09-06T22:20  &  X2.1   &   11283 &  E & 1.46 & 0.66   & 1.50   & 1.59   & 0.43      & 1.06\\
    8   & SOL2011-10-02T00:50  & M3.9   &   11305 &  C & 0.84 & 0.48   & 0.64   & 0.69   & -0.66     & -0.58\\
    9   & SOL2012-01-23T03:59  & M8.7   &   11402 &  E & 2.44 & 1.60   & 3.59   & 3.79   & -1.18     & -1.77\\
    10 & SOL2012-03-07T00:24  & X5.4   &    11429 &  E & 2.71 & 2.85   & 9.25   &  9.73   & -6.36    & -10.70\\
    11 & SOL2012-03-09T03:53  & M6.3  &    11429 &  E & 2.39 & 2.11   & 6.26   & 6.62   & -3.06     & -7.80\\
    12 & SOL2012-05-10T04:18  & M5.7  &    11476 &  C & 3.21 & 2.85   & 3.40   & 3.65   & 2.65      & 3.24\\
    13 & SOL2012-07-02T10:52  & M5.6  &    11515 &  E & 2.09 & 1.40   & 1.36   & 1.45   & -1.48     & -1.07\\
    14 & SOL2012-07-05T11:44  & M6.1  &    11515 &  C & 3.20 & 2.65   & 5.35   & 5.58   & -4.89    & -8.13\\
    15 & SOL2012-07-12T16:49  & X1.4  &     11520 &  E & 4.47 & 3.77   & 8.12   & 8.53   & 7.63     & 13.50\\
    16 & SOL2013-04-11T07:16  & M6.5 &     11719 &  E & 1.23 & 0.47   & 0.74   & 0.79   & 0.15     & 0.38\\
    17 & SOL2013-10-24T00:30  & M9.3 &     11877 &  E & 2.48 & 1.62   & 1.87   & 2.00   &1.25      & 0.90\\
    18 & SOL2013-11-01T19:53  & M6.3 &     11884 &  C & 1.83 & 0.99   & 1.26   & 1.34   & 0.11     & 0.30\\
    19 & SOL2013-11-03T05:22  & M4.9 &     11884 &  C & 1.73 & 0.83   & 0.89   & 0.97   & 0.19     & 0.18\\
    20 & SOL2013-11-05T22:12  & X3.3 &      11890 &  E & 4.27 & 3.90   & 5.51   & 5.82   & 2.62     & 4.07\\
    21 & SOL2013-11-08T04:26  & X1.1 &      11890 &  E & 2.64 & 2.18   & 2.30   & 2.43   & 1.69     & 2.11\\
    22 & SOL2013-12-31T21:58  & M6.4 &     11936 &  E & 2.31 & 1.32   & 3.06   & 3.19   & -0.20    & -1.62\\
    23 & SOL2014-01-07T10:13  & M7.2 &     11944 &  C & 4.68 & 5.99   & 6.82   & 7.16   & 12.25   & 11.90\\
    24 & SOL2014-01-07T18:32  & X1.2 &      11944 &  E & 5.21 & 5.97   & 6.32   & 6.67   & 11.25   & 10.70\\
    25 & SOL2014-02-02T09:31  & M4.4 &     11967 &  C & 4.11 & 3.68   & 7.16   & 7.70   & -1.60   & -4.94\\
    26 & SOL2014-02-04T04:00  & M5.2 &     11967 &  C & 4.15 & 3.93   & 7.36   & 7.83   & -2.61   & -7.75\\
    27 & SOL2014-03-29T17:48  & X1.1 &      12017 &  E & 1.31 & 0.48   & 1.44   & 1.51   & 1.09    & 2.43\\
    28 & SOL2014-04-18T13:03  & M7.3 &     12036 &  E &  2.19 & 1.07   & 2.42   & 2.54   & 1.61   & 2.92\\
    29 & SOL2014-09-10T17:45  & X1.6 &      12158 &  E & 1.55 & 1.24   & 2.07   & 2.20   & -1.98   & -2.44\\
    30 & SOL2014-09-28T02:58  & M5.1 &     12173 &  E & 3.58 & 2.90   & 6.92   & 7.12   & -3.46   & -10.90\\
    31 & SOL2014-10-22T14:28  & X1.6 &      12192 &  C & 7.79 & 13.16 & 16.10 & 16.60 & -31.54 & -24.60\\
    32 & SOL2014-10-24T21:41  & X3.1 &      12192 &  C & 9.09 & 16.89 & 22.90 & 23.50 & -46.57 & -35.90\\
    33 & SOL2014-11-07T17:26  & X1.6 &      12205 &  E & 2.83 & 1.94   & 8.87   & 9.16   & -1.04   & 4.45\\
    34 & SOL2014-12-04T18:25  & M6.1 &     12222 &  C & 2.26 & 1.89   & 1.36   & 1.46   & -1.30   & -0.99\\
    35 & SOL2014-12-17T04:51  & M8.7 &     12242 &  E & 2.80 & 1.86   & 3.20   & 3.35   & 2.48     & 3.70\\
    36 & SOL2014-12-18T21:58  & M6.9 &     12241 &  E & 1.83 & 1.55   & 1.84   & 1.98   & 1.23     & 1.17\\
    37 & SOL2014-12-20T00:28  & X1.8 &      12242 &  E & 4.12 & 3.31   & 6.47   & 6.83   & 5.30     & 6.63\\
    38 & SOL2015-03-11T16:21  & X2.1 &      12297 &  E & 1.81 & 1.21   & 4.51   & 4.72   & 1.61     & 4.17\\
    39 & SOL2015-03-12T14:08  & M4.2 &     12297 &  C & 1.65 & 1.02   & 2.77   & 2.96   & 0.79     & 1.88\\
    40 & SOL2015-06-22T18:23  & M6.5 &     12371 &  E & 2.39 & 2.62   & 7.07   & 7.32   & -5.88    & -12.50\\
    41 & SOL2015-06-25T08:16  & M7.9 &     12371 &  E & 3.13 & 2.74   & 7.20   & 7.44   & -5.85    & -12.20\\
    42 & SOL2015-08-24T07:33  & M5.6 &     12403 &  C & 3.27 & 3.48   & 3.13   & 3.30   & 0.25     & 0.51\\
    43 & SOL2015-09-28T14:58  & M7.6 &     12422 &  C & 2.38 & 2.16   & 1.59   & 1.71   & -1.05    & -0.82\\
    44 & SOL2017-09-04T20:33  & M5.5 &     12673 &  E & 1.82 & 1.57   & 6.13   & 6.37   & -3.74    & -10.10\\
    45 & SOL2017-09-06T12:02  & X9.3 &      12673 &  E & 2.66 & 2.61   & 10.80 & 11.30 & -5.10    & -15.60\\
    \hline
  \end{tabular}

 \textbf{Notes.} \textbf{$\phi^{\prime}$} is half of the total unsigned flux; \textbf{$E_T$} is the total magnetic field; \textbf{$E_{F1}$} and \textbf{$E_{F2}$} are the free magnetic energies calculated with different methods; \textbf{$H_R$} is the relative magnetic helicity and \textbf{$H_J$} is the non-potential helicity.
  \tablenotetext{a}{E--eruptive, C--confined.}
  \label{tab:event_list}
\end{table*}

The computations of the magnetic energy and helicity are based on the coronal magnetic field extrapolations as already carried out in our previous study~\citep{DuanA2019}. The 3D pre-flare coronal field were reconstructed from the {\SDO} Helioseismic and Magnetic Imager~\citep[HMI;][]{Hoeksema2014} vector magnetograms using the CESE--MHD--NLFFF code~\citep{JiangC2013NLFFF}. For the vector magnetograms,  we used the data product of the Space-weather HMI Active Region Patch~\citep[SHARP;][]{Bobra2014}, in which the $180^{\circ}$ ambiguity has been resolved by using the minimum energy method, the coordinate system has been modified via the Lambert method, and the projection effect has been corrected. In order to avoid the possible artifacts introduced by the strong flare emission, the last available magnetogram for at least 10 minutes before the flare start time was utilized. The vector magnetograms are preprocessed before being put into the extrapolation code in order to reduce the data noise as well as the Lorentz force contained in the photosphere~\citep{Jiang2014Prep}. All the extrapolations are performed with spatial resolution of 1~arcsec, and the field of view (FoV) of extrapolation volumes are mostly consistent with the FoV of the SHARP data, except for those containing more than one AR, for which we cut off the unrelated ARs. For each extrapolation, the height of the volume is chosen to be equal to the smaller one of the two horizontal lengths.

For each event, we calculated the total unsigned flux $\phi$, the total ($E_T$), potential ($E_P$) and free ($E_F$) magnetic energy, as well as the relative ($H_R$) and non-potential ($H_{J}$) helicities, for which the expressions are given below.

The free magnetic energy represents the deviation of the total magnetic energy from potential-field energy. Since the computation of free energy is influenced by numerical error of $\nabla\cdot\vec B=0$~\citep{Valori2013}, and a considerable large divergence error might  render the computation of free energy totally unreliable (for example, resulting in a negative value). Therefore,  here we calculated it with two different ways to estimate how much this effect impacts the computation of the free energy in our extrapolated NLFFFs. In the first approach, we calculated the total magnetic energy $E_T$ and the corresponding potential energy $E_P$ first, and then get the free energy $E_{F1}$ as the difference between $E_T$ and $E_P$, namely,
\begin{equation}\label{eq1}
  E_{F1} = E_T - E_P = \frac{1}{8\pi}\int_V{B^2}dV -  \frac{1}{8\pi}\int_V{B_P^2}dV,
\end{equation}
here $\vec B$ is the magnetic field from NLFFF extrapolation and can be treated as the total magnetic field; $\vec B_P$ is the potential field sharing the normal component with $\vec B$ on the boundary of $V$; In the second approach, the ``free field'' or the current-carrying field $\vec B_J$ was firstly computed as $\vec B_J = \vec B - \vec B_P$, and then the free energy can be calculated as
\begin{equation}\label{eq2}
  E_{F2} = \frac{1}{8\pi} \int_V B_{J}^2 dV.
\end{equation}
If the extrapolated field $\vec B$ is perfectly divergence-free, the two ways of computing free energy will lead to the same result. Thus, the difference of these two ways of free energy calculation reflects the divergence errors. 

The relative magnetic helicity for a 3D magnetic field $\vec B$ in a finite volume $V$ is computed as
\begin{equation}\label{eq3}
  H_R = \int_V (\vec A+\vec{A_P})\cdot (\vec B-\vec{B_P}) dV,
\end{equation}
where $\vec A$ and $\vec {A_P}$ are the vector potentials of the two magnetic field, satisfying $\vec B = \nabla \times \vec A$ and $\vec B_P = \nabla \times \vec A_P$, respectively. Here we followed the procedure as introduced by~\citet{Valori2012B} to calculate the vector potentials.

Furthermore, the non-potential (i.e., current-carrying) helicity is simply defined as
\begin{equation}\label{eq4}
  H_J = \int_V \vec{A_J}\cdot \vec{B_J} dV,
\end{equation}
where $\vec{A_J} = \vec A - \vec{A_P}$ and thus $\vec{B_J}=\nabla \times \vec A_J$; and the volume-threading helicity between $\vec B_P$ and $\vec B$ is computed as
\begin{equation}\label{eq5}
  H_{PJ} = 2\int_V\vec{A_P}\cdot\vec{B_J} dV,
\end{equation}
which can be derived from $H_R = H_J + H_{PJ}$.

\begin{figure}
  \centering
  \includegraphics[width=0.45\textwidth]{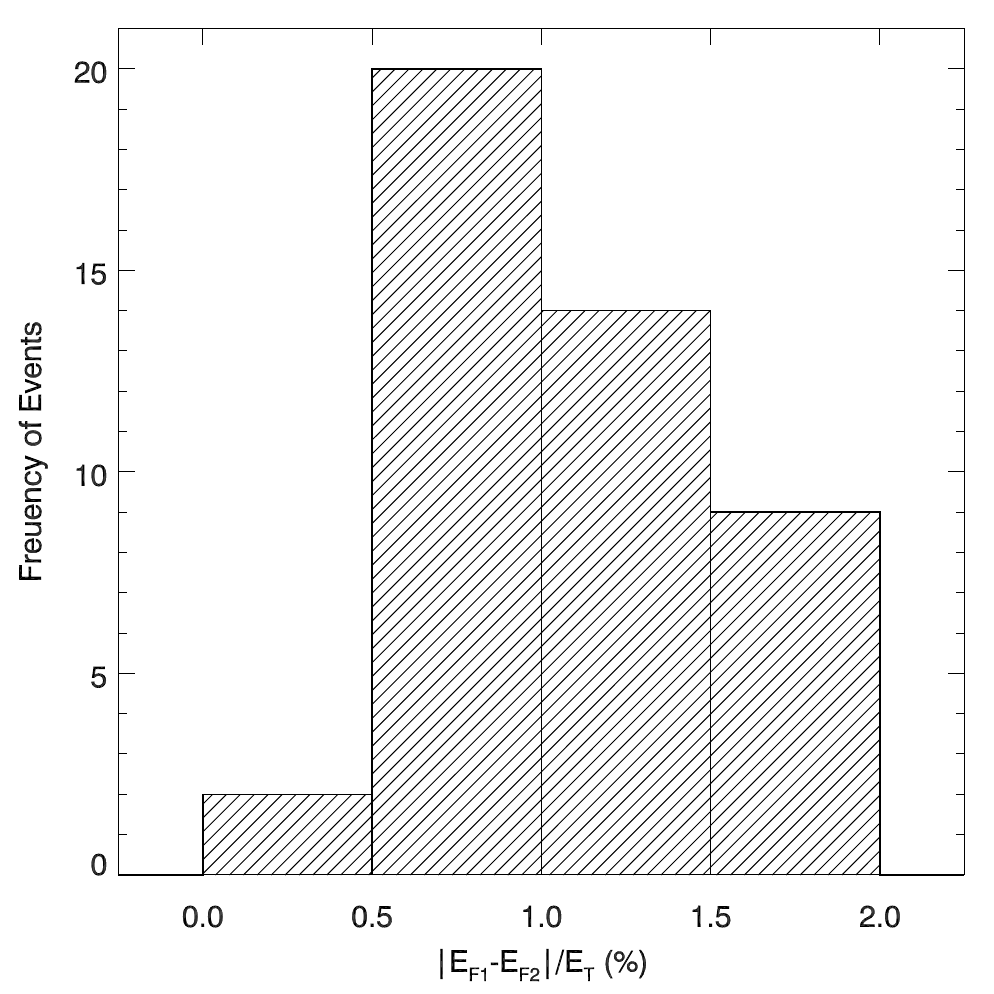}
  \caption{{Distribution of the discrepancy between $E_{F1}$ and $E_{F2}$ as normalized by the corresponding $E_T$ for all the events.}}
  \label{f5}
\end{figure}

In Table~\ref{tab:event_list}, we listed the calculated parameters of $E_T$, $E_{F1}$, $E_{F2}$, $H_R$, and $H_J$. As can be seen in~\Fig~\ref{f5}, we found the discrepancy between $E_{F1}$ and $E_{F2}$ is less than 2\% of the corresponding $E_T$, which indicates that the influence of the divergence error is very small. Therefore, we use $E_F = E_{F1}$ in all the analysis in the next section. With all the extensive parameters obtained, subsequently we computed the intensive quantities such as the energy ratio $E_F/E_P$, the helicity ratio $|H_J/H_R|$, and the normalized helicities, $|H_R/\phi^{'2}|$ and $|H_J/\phi^{\prime2}|$, where $\phi^{\prime}$ is the half of the total unsigned flux which has $\phi^{\prime}=\frac{1}{2}\phi=\frac{1}{2}\int_{S(z=0)}|B_z|dS$.

\section{Results}
\label{sec:res}

\begin{figure*}
  \centering
  \includegraphics[width=0.7\textwidth]{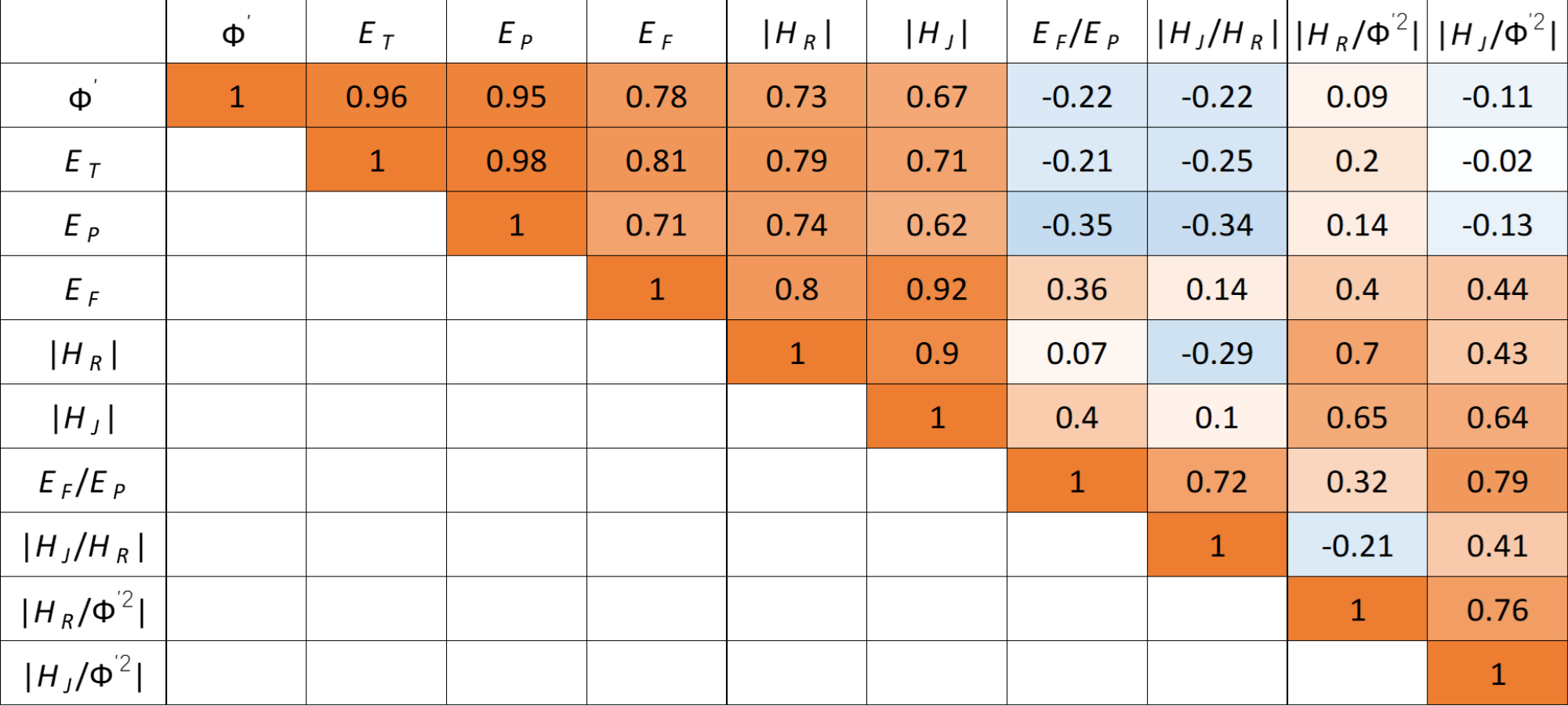}
  \caption{{The Spearman rank correlation between different parameters. Colors correspond to the strength of the correlation coefficient, $r_s$, between each variable pair. Orange (blue) represents a positive (negative) $r_s$, and the deeper the color, the higher the absolute value of $r_s$.}}
  \label{f6}
\end{figure*}

Firstly, we are interested in how the parameters are correlated with one another. \Fig~\ref{f6} shows a matrix of the correlation coefficients (CCs) between all the parameters. Note that here we used the Spearman rank correlation $r_s$ because the Pearson CC is used to measure the linear relationship between variables and therefore is not optimal for nonlinearly related variables. The Spearman rank correlation provides a measure of the monotonic relationship between variables and is thus more suitable for this study. Following~\citet{Kazachenko2017}, we describe the qualitative strength of the correlation using the absolute value of $ r_s \in [0.2, 0.39]$--weak, $r_s \in [0.4, 0.59]$--moderate, $r_s \in [0.6, 0.79]$--strong, and $r_s \in [0.8, 1.0]$--very strong.

As can be seen, all the extensive parameters (i.e., the first 6 ones) are strongly correlated with one another with $r_s$ all above $0.6$. Among them, the total magnetic energy has the highest $r_s$ of $0.98$ with the potential energy, which agrees well with a large-sample statistical study based on different NLFFF extrapolation codes~\citep{Aschwanden2014a}, though the CC is not explicitly given in that study. Both the two energies $E_T$ and $E_P$ have very strong correlation with the unsigned flux $\phi^{\prime}$ with $r_s \ge 0.95$, which is easy to understand since more flux can hold more global energies. The free magnetic energy $E_F$ has also strong correlation with the three parameters $E_T$, $E_P$, and $\phi^{\prime}$ (the highest $r_s = 0.81$ with $E_T$, which is also suggested in~\citet{Aschwanden2014a}). The relative helicity $H_R$ is also similarly correlated with the flux and energies, but has the strongest correlation with the free energy ($r_s = 0.80$). Such a very strong correlation between $E_F$ and $H_R$ is previously found by~\citet{Tziotziou2012} based on an investigation of 42 different ARs (with either flaring or non-flaring ones), who demonstrated that there is a statistically monotonic correlation between the free energy and relative helicity. The correlation of the non-potential helicity $H_J$ with other parameters is investigated for the first time here. $H_J$ is strongly correlated with $H_R$, and the former is overall somewhat less correlated with $\phi^{\prime}$, $E_T$ and $E_P$ than as the latter. Notably, $H_J$ is the most highly correlated with the free energy with $r_s = 0.92$, which is also the highest CC of the free energy with all other parameters. But this is not unexpected since according to their definitions (see~\Eqs~\ref{eq2} and~\ref{eq4}), they are both based solely on the current-carrying field $\vec B_J$ which is a self-closed magnetic field and thus with weaker correlation to the potential field (and the unsigned flux) than as the relative helicity.

The intensive parameters quite interestingly show more negative than positive correlation with the first three extensive parameters. For instance, the non-potentiality energy ratio $E_F/E_P$, the helicity ratio $|H_J/H_R|$, and the normalized non-potential helicity $|H_J/\phi^{\prime2}|$, are all negatively correlated with the total unsigned flux, the potential energy, and the total energy, though the correlations are rather weak (with CCs all below $0.4$). This hints that when ready for producing major flares, the larger ARs (i.e., with more flux and thus more global energies) are less deviated from the potential state than the smaller ARs. When focusing on the relationship between the intensive parameters themselves, we find that the energy ratio $E_F/E_P$ is strongly correlated with both the helicity ratio $|H_J/H_R|$ and the normalized non-potential helicity $|H_J/\phi^{\prime2}|$, especially for latter one the CC reaching the highest value $0.79$ of all, and this CC is also the highest one of $|H_J/\phi^{\prime2}|$ with all others. This is consistent with the strong correlation between the two extensive parameters $E_F$ and $H_J$ and since the denominators $E_P$ and $\phi^{\prime2}$ are also highly correlated. On the other hand, the normalized total helicity $|H_R/\phi^{\prime2}|$ is less correlated with the other intensive parameters, suggesting its weaker sensitivity (than the others) on the degree of the non-potentiality (and complexity) of the magnetic field.

\begin{figure*}
  \centering
  \includegraphics[width=0.8\textwidth]{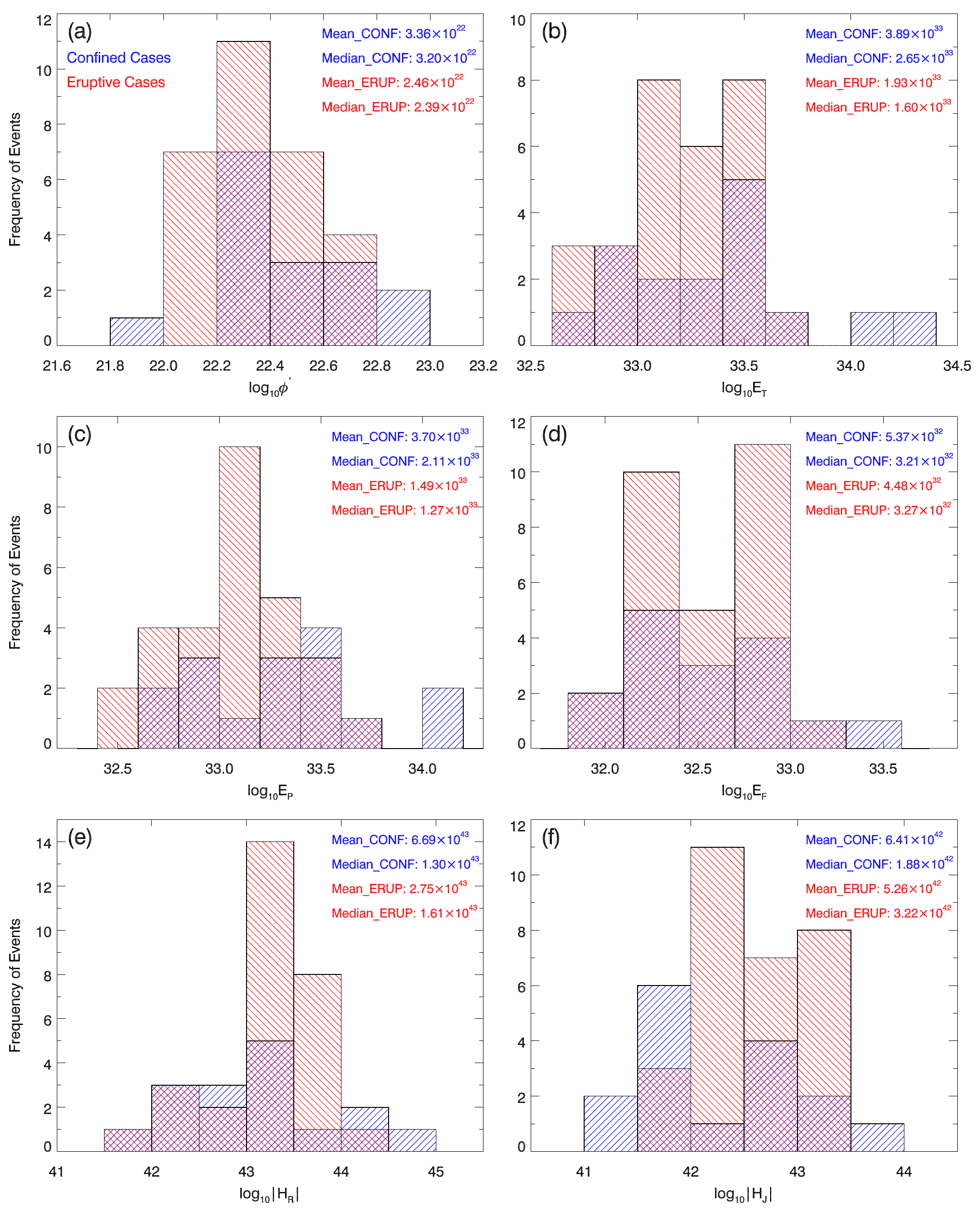}
  \caption{{Histograms of the extensive-type indices. (a) Half of the total unsigned flux $\phi^{\prime}$; (b) the total magnetic energy $E_T$; (c) the magnetic potential energy $E_P$; (d) the magnetic free energy $E_F$; (e) the absolute value of the relative magnetic helicity $|H_R|$ and (f) the absolute value of the current-carrying helicity $|H_J|$.}}
  \label{f1}
\end{figure*}

Next we explore whether the parameters behave differently in the two types of flares, and whether any of them stand out for discriminating the different types, which is the main purpose of this study. In~\Fig~\ref{f1}, we show distributions of all the extensive parameters for the two types of flare events separately, including (a) the unsigned flux $\phi^{\prime}$, (b) the total magnetic energy $E_T$, (c) the magnetic potential energy $E_P$ and (d) free energy $E_F$, as well as (e) the relative magnetic helicity $H_R$, and (f) the non-potential magnetic helicity $H_J$. The eruptive flares are colored in red, and the confined ones in blue. Meanwhile, the mean and median values of each parameter for both types of flares are presented on each panel. We find that most of the flares originate from ARs with total unsigned flux on the order of $10^{22}$~Mx. The average and median $\phi^{\prime}$ for eruptive flares are $2.46\times10^{22}$~Mx and $2.39\times10^{22}$~Mx, while for confined cases they are $3.36\times10^{22}$~Mx and $3.20\times10^{22}$~Mx, which are larger than the values for eruptive ones. This is consistent with a large-sample statistical study by~\citet{LiT2021}, in which they found that flares of the same GOES class but originating from an AR of larger magnetic flux are much more likely to be confined.

\Fig~\ref{f1} (b) shows that the total magnetic energy $E_T$ mostly distribute around $10^{33}$~erg for both confined and eruptive flares. Similar to the distribution of magnetic flux, both the average and median $E_T$ for confined flares ($3.89\times10^{33}$~erg and $2.65\times10^{33}$~erg) are larger than the corresponding values of the eruptive ones (with $1.93\times10^{33}$~erg and $1.60\times10^{33}$~erg).

The potential energy $E_P$ (\Fig~\ref{f1}c) mostly distributes around $10^{33}$~erg, while the free energy $E_F$ (\Fig~\ref{f1}d) around $10^{32}$~erg, which is smaller than that of the potential energy by roughly an order of magnitude.  Both the average ($3.70\times10^{33}$~erg) and median ($2.11\times10^{33}$~erg) of $E_P$ for confined flares are larger than values of the eruptive ones ($1.49\times10^{33}$~erg and $1.27\times10^{33}$~erg). For $E_F$, the average (median) value of the confined flares is $5.37\times10^{32}$~erg ($3.21\times10^{32}$~erg), and it is larger (slightly smaller) than the corresponding values of the eruptive cases, which is $4.48\times10^{32}$~erg and $3.27\times10^{32}$~erg, respectively.

\begin{figure*}
  \centering
  \includegraphics[width=0.68\textwidth]{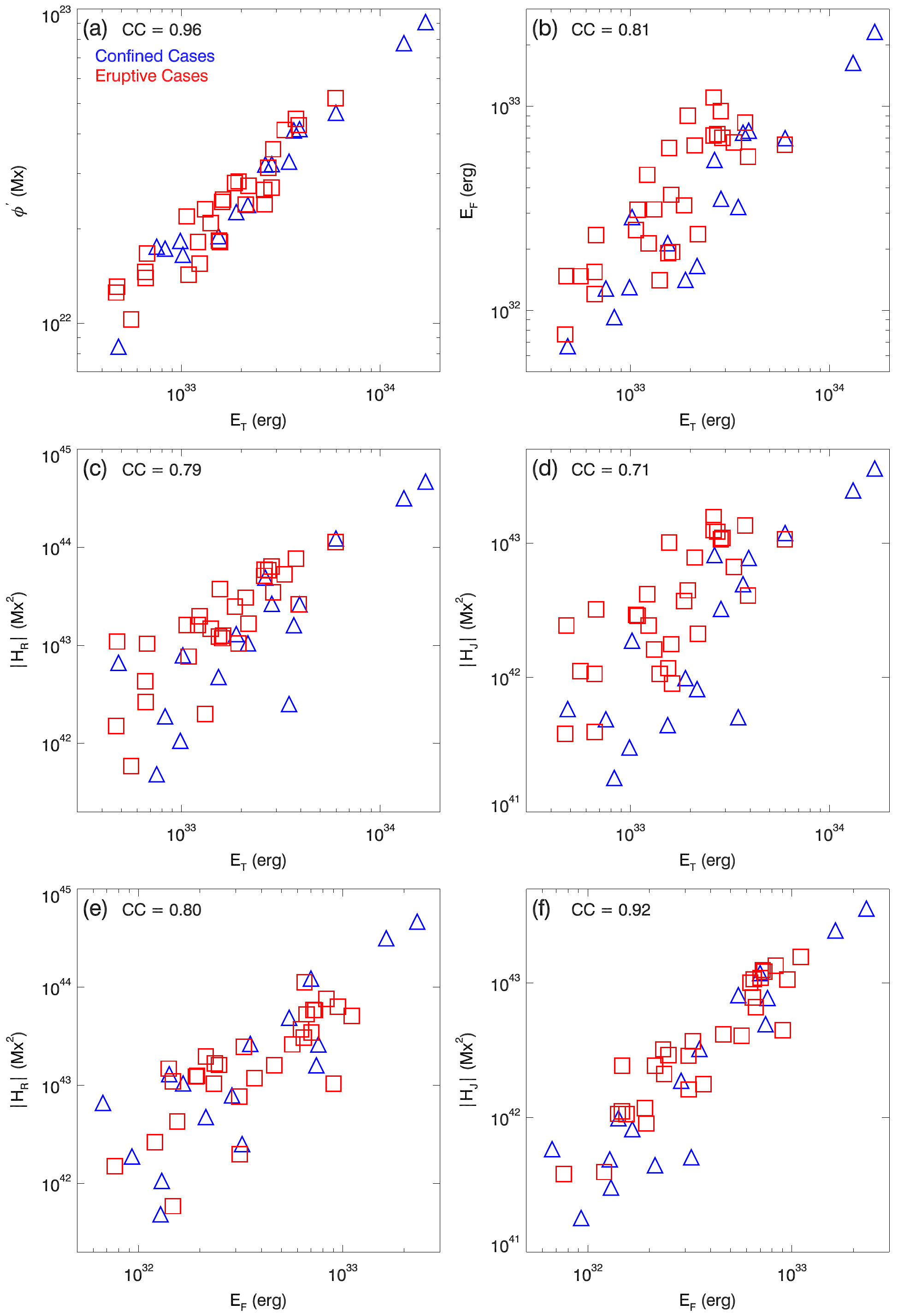}
  \caption{{Scatter diagrams of $E_T$ vs. $\phi^{\prime}$, $E_T$ vs. $E_F$, $E_T$ vs. $|H_R|$, $E_T$ vs. $|H_J|$, $E_F$ vs. $|H_R|$ and $E_F$ vs. $|H_J|$ for all events in (a)-(f), respectively. The red boxes (blue triangles) denote eruptive (confined) flares, and their Spearman CCs are labeled on the panels.}}
  \label{f4}
\end{figure*}

\Fig~\ref{f1} (e) shows that for confined flares, the relative magnetic helicity $|H_R|$ has a broad distribution of $10^{42}\sim10^{44}$~Mx$^2$. Its average value ($6.69\times10^{43}$~Mx$^2$) is clearly larger than the median value ($1.30\times10^{43}$~Mx$^2$) since 3 cases (out of 16) have $|H_R|$ larger than $10^{44}$~Mx$^2$. For eruptive flares, most of them (22 out of 29) have $|H_R|$ between $10^{43}\sim10^{44}$~Mx$^2$, while only one of them has $|H_R|$ larger than $10^{44}$~Mx$^2$;  and the average and median values are $2.75\times10^{43}$~Mx$^2$ and $1.61\times10^{43}$~Mx$^2$, respectively. The confined cases have larger average $|H_R|$ than eruptive ones, while the eruptive flares have slightly larger median $|H_R|$ than the confined one.

\Fig~\ref{f1} (f) shows for the non-potential helicity $|H_J|$, the average and median $|H_J|$ are $6.41\times10^{42}$~Mx$^2$ and $1.88\times10^{42}$~Mx$^2$ for confined cases and $5.26\times10^{42}$~Mx$^2$ and $3.22\times10^{42}$~Mx$^2$ for the eruptive ones. Half of (8 out of 16) the confined flares have $|H_J|$ from $10^{41}\sim10^{42}$~Mx$^2$ and the other half ranges from $10^{42}\sim10^{43}$~Mx$^2$. Comparatively, the distribution of $|H_J|$ for eruptive flares are slightly more concentrated.

\Fig~\ref{f4} shows the scatter diagrams in different two groups of the different extensive parameters. These diagrams also indicate clearly the correlations of the parameters. However, the parameters for the two types of events (i.e., eruptive and confined flares) distribute almost evenly in the diagrams albeit that they have larger average values for confined flares than for eruptive flares, thus overall the extensive parameters show no apparent systematic difference between the eruptive and confined flares and cannot discriminate the two types. We note that 3 confined cases are produced by AR 11944 (on 2014-01-07T10:13) and AR 12192 (on 2014-10-22T14:28 and 2014-10-24T21:41), respectively. Both the two ARs have very large size of areas of magnetic field concentration, especially, the AR 12192 hosts the largest sunspot group in solar cycle 24~\citep{SunXD2015}, which thus leads to a great possession of the extensive parameters.

\begin{figure*}
  \centering
  \includegraphics[width=0.8\textwidth]{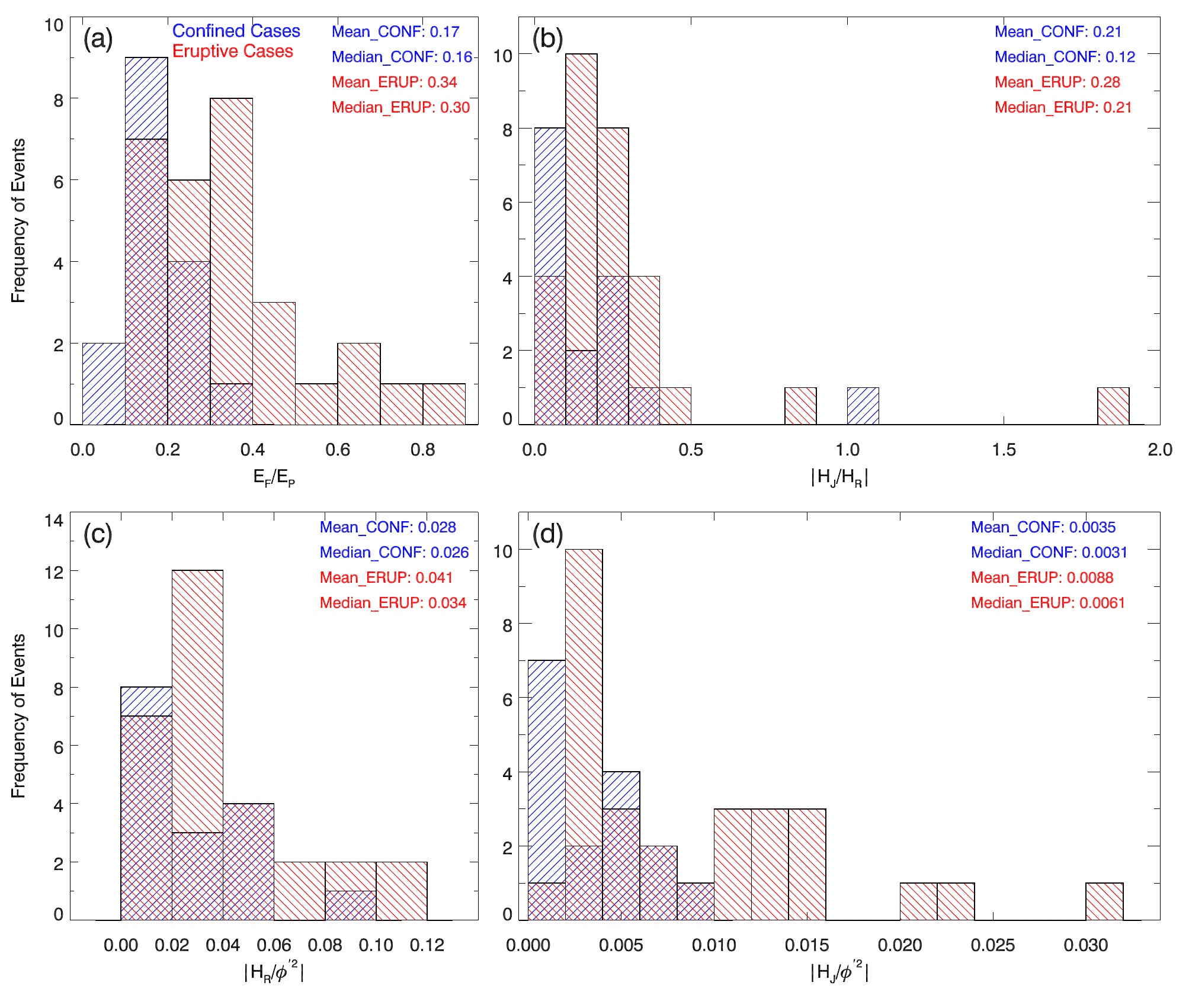}
  \caption{{Histograms of the intensive-type indices. (a) The ratio of the magnetic free energy to potential energy $E_F/E_P$; (b) the ratio of current-carrying helicity to the relative magnetic helicity $|H_J/H_R|$; (c) $|H_R/\phi^{\prime2}|$, here $\phi^{\prime}$ is the half of the total unsigned magnetic flux; and (d) $|H_J/\phi^{\prime2}|$.}}
  \label{f2}
\end{figure*}

\Fig~\ref{f2} shows the histograms of the intensive parameters for the two types of flares, including $E_F/E_P$, $|H_J/H_R|$, $|H_R/\phi^{\prime2}|$, as well as $|H_J/\phi^{\prime2}|$. As can be seen for all intensive parameters, both their average and median values for eruptive flares are significantly larger than the corresponding values for confined flares, especially the energy ratio $E_F/E_P$ and the normalized non-potential helicity $|H_J/\phi^{\prime2}|$, for both of which the values of the eruptive flares are around double of those of the confined ones. This is in contrast to the result for the extensive parameters, and indicates that the field for eruptive flares have overall high degrees of non-potentiality (and complexity) than that of the confined flares. Furthermore, the distribution of $E_F/E_P$ shows apparent difference between the two types of flares. All the confined cases have $E_F/E_P\le$ 0.4, while 27$\%$ (8 out of 29) of the eruptive cases have $E_F/E_P\ge$ 0.4 (with the largest value of over 0.8). This difference can be similarly seen in the distribution of $|H_J/\phi^{\prime2}|$, as all the confined events have value below 0.01 while over 20\% (6 in 29) of the eruptive events above 0.01. In \Fig~\ref{f2} (b), there are two cases with $|H_J/H_R|$ larger than 1, and we found in those cases $H_R$ and $H_J$ is positive, while $H_{PJ}$ is negative.

\begin{figure*}
  \centering
  \includegraphics[width=0.86\textwidth]{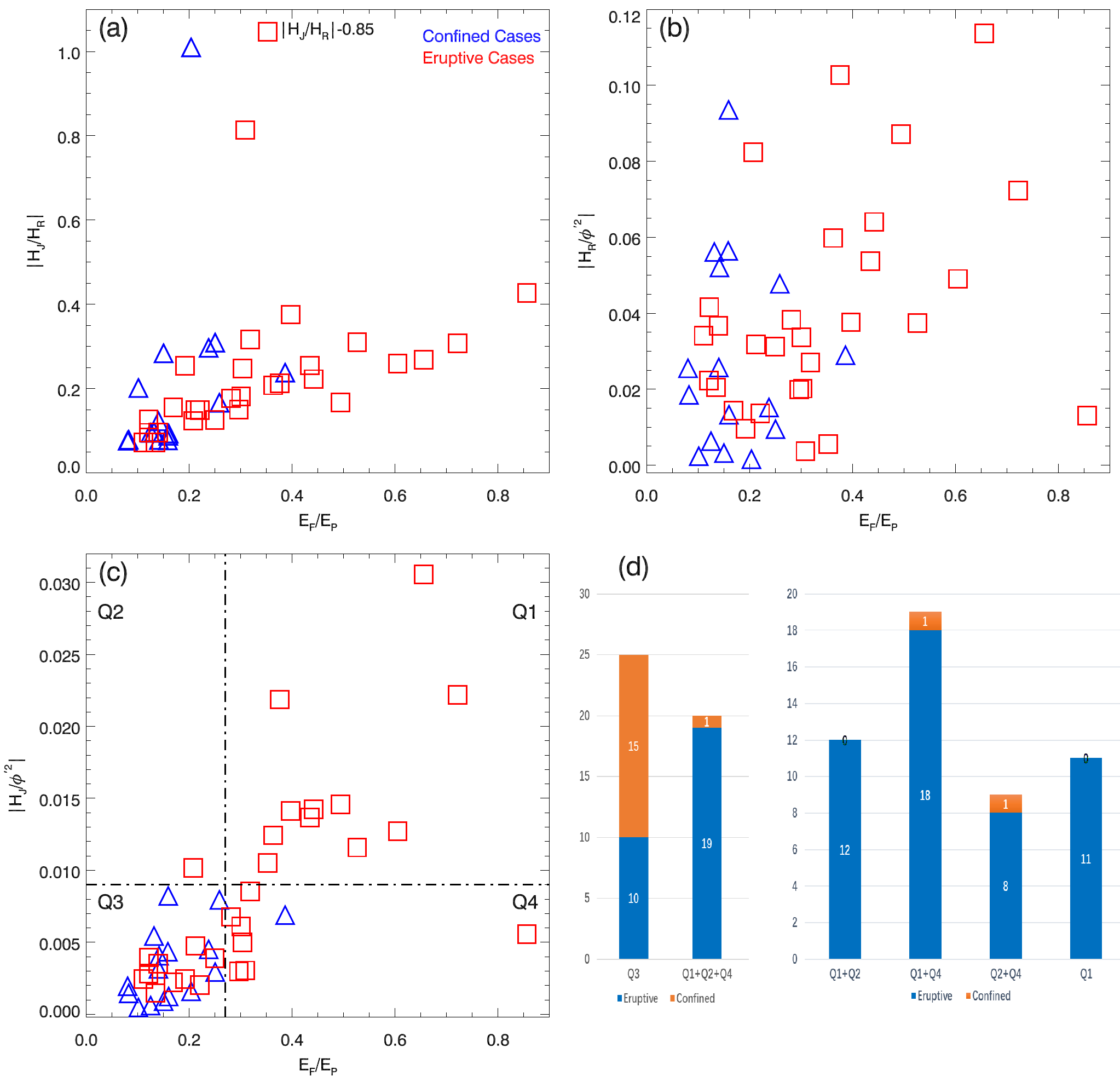}
  \caption{{Scatter diagrams of $E_F/E_P$ vs. $|H_J/H_R|$, $E_F/E_P$ vs. $|H_R/\phi^{\prime2}|$, and $E_F/E_P$ vs. $|H_J/\phi^{\prime2}|$ for all events in (a)-(c), respectively. The red boxes (blue triangles) denote eruptive (confined) flares. Note that the maximum value of $|H_J/H_R|$ is 1.89, and for a better display of the plot, we show it as ($|H_J/H_R|$-0.85) in (a). The vertical dashed line in panel (c) shows $E_f/E_p=0.27$, and the horizontal dashed line shows $|H_J/\phi^{\prime2}|=0.009$. Based on the lines in panel (c), the distribution of all the events can be divided into four quadrants, Q1, Q2, Q3, and Q4. (d) Histograms for numbers of events in Q3, Q1+Q2+Q4, Q1+Q2, Q1+Q4, Q2+Q4, and Q1.}}
  \label{f3}
\end{figure*}

In \Fig~\ref{f3} (a)-(c), we show the scatter diagrams of $|H_J/H_R|$ versus $E_F/E_P$,  $|H_R/\phi^{\prime2}|$ versus $E_F/E_P$ and $|H_J/\phi^{\prime2}|$ versus $E_F/E_P$ for the events. These plots show systematic difference between the two types of events which is not seen in the extensive parameters (\Fig~\ref{f4}). In particular, from the distribution of eruptive and confined flares in the parameter spaces, it is possible to empirically identify critical values that can discriminate a majority of the different events. Panel (a) shows that most of the events have $|H_J/H_R|\le 0.5$ and it cannot discriminate eruptive flares from confined ones effectively. Panel (b) shows that $|H_R/\phi^{\prime2}|$ also has poor ability to distinguish different types of flares. On the other hand, as marked on panel (c), when we use critical values of $|H_J/\phi^{\prime2}|_\texttt{crit}=0.009$ and $(E_F/E_P)_\texttt{crit}=0.27$, respectively, the distribution of the events falls into four quadrants which are defined as Q1 ($|H_J/\phi^{\prime2}|\ge0.009$ and $E_F/E_P\ge0.27$), Q2 ($|H_J/\phi^{\prime2}|\ge0.009$ and $E_F/E_P<0.27$), Q3 ($|H_J/\phi^{\prime2}|<0.009$ and $E_F/E_P<0.27$), and Q4 ($|H_J/\phi^{\prime2}|<0.009$ and $E_F/E_P\ge0.27$). The histograms for events falling into different quadrants are shown in \Fig~\ref{f3} (d). As can be seen, all the events with $|H_J/\phi^{\prime2}|$ above $|H_J/\phi^{\prime2}|_\texttt{crit}$ (i.e., in Q1+Q2) erupted. Thus $|H_J/\phi^{\prime2}|>|H_J/\phi^{\prime2}|_\texttt{crit}$ can be regarded as a sufficient condition for an eruptive flare. For all the events with $E_F/E_P$ above $(E_F/E_P)_\texttt{crit}$ (i.e., Q1+Q4), 95\% erupted (18 in 19). Over 94\% of confined events (15 in 16) reside in Q3. If doing a prediction for eruptive or confined events in all the 45 events using these critical values, over 75\% are successfully predicted, and the remaining 11 events include 10 eruptive ones in Q3 and 1 confined one in Q4.

\section{Summary}
\label{sec:con}
In this paper, we carried out a survey of important parameters related to magnetic energy and helicity for 45 major solar flares (generally above GOES M5 class), with the main purpose to explore whether there are parameters that can effectively discriminate the eruptive and confined flares. These investigated parameters include the extensive-type quantities which are, respectively, the unsigned flux $\phi^{\prime}$, the total magnetic energy $E_T$, potential magnetic energy $E_P$, free magnetic energy $E_F$, the relative magnetic helicity $H_R$, as well as the non-potential (current-carrying) magnetic helicity $H_J$, and the intensive-type indices (mainly as the ratios of different extensive indices, e.g., $E_F/E_P$, $|H_J/H_R|$, $|H_R/\phi^{\prime2}|$ and $|H_J/\phi^{\prime2}|$). Using the CESE--MHD--NLFFF method with {\SDO}/HMI vector magnetograms as input, we reconstructed the coronal magnetic fields immediately prior to the flares for all events, and then calculated the parameters. The results are summarized as following.

(1) All the extensive parameters are strongly correlated between one another with Spearman CCs all above 0.6. Importantly, both the relative helicity $H_R$ and the non-potential helicity $H_J$ have the strongest correlation with the free energy (with $r_s = 0.8$ and $r_s = 0.92$, respectively) than with all other parameters, and in turn the correlation of the free energy with the non-potential helicity $H_J$ is the highest than with all other parameters. This confirms the intrinsic relationship between the non-potentiality and complexity of the coronal magnetic field.

(2) The intensive parameters show negative (but rather weak) correlation with the first three extensive parameters, which hints that when ready for producing major flares, the larger ARs (i.e., with more flux and thus more global energies) are less deviated from the potential state than the smaller ARs. The energy ratio $E_F/E_P$ (which measure the degree of non-potentiality) is strongly correlated with both the normalized non-potential helicity $|H_J/\phi^{\prime2}|$ and the helicity ratio $|H_J/H_R|$. Especially, $E_F/E_P$ and $|H_J/\phi^{\prime2}|$ has the strongest correlation among all others with these two parameters.

(3) On average, all the extensive parameters have larger average values for confined flares than for eruptive flares (since the confined events have overall larger magnetic flux), but their distributions show no apparent systematic difference between the eruptive and confined events. Therefore, in general, all the extensive quantities cannot distinguish the eruptive and confined potential of the flare-producing AR.

(4) In contrast to the extensive parameters, all the intensive parameters have significantly larger average and median values for eruptive flares than the confined events. This indicates that the field for eruptive flares have overall high degrees of non-potentiality and complexity than that of the confined flares. In distinguishing the two types of events, the intensive parameters show certain ability that is absent for the extensive ones. Among them, the energy ratio $E_F/E_P$ and the normalized current-carrying helicity $|H_J/\phi^{\prime2}|$, are the highest capable of distinguishing the pre-flare corona of ARs that possibly produce a large eruptive or confined flare. Particularly, by using the characteristic pre-flare values of $E_F/E_P\ge0.27$ and $|H_J/\phi^{\prime2}|\ge0.009$ over 75\% of the events are successfully discriminated between eruptive and confined flares. On the other hand, $|H_R/\phi^{\prime2}|$ and $|H_J/H_R|$ are less important in differentiating the eruptive flares from the confined ones. Our result partially supports the finding of~\citet{Gupta2021}, who showed that the events with $E_F/E_T>0.2$ (which is equivalent to $E_F/E_P>0.25$, a value very close to our derived one of 0.27) and $|H_J/\phi^{\prime2}|>0.005$ is more likely to produce eruptive flares. However, we do not find systematic difference in the value of $|H_J/H_R|$ between the two types of flares, which is at variance with~\citet{Gupta2021}, where they found that $|H_J/H_R|$ is distinctly different for ARs producing different types of flares.

Finally it is interesting to compare the result of this work with our previous work~\citep{DuanA2019, DuanA2021apjl} which are based on the same data set but focused on the special magnetic configuration, i.e.,  the magnetic field rope (MFR), and the related ideal MHD instabilities (namely the torus instability and kink instability) in determining the eruptive and confined types of the flares. In that work, by a comprehensive analysis of the two control parameters, which are decay index of the strapping field ($n$) for torus instability and the maximum twist number ($T_w$) in the MFR for kink instability, we found two critical values of $n_\texttt{crit} = 1.3$ and $|T_w|_\texttt{crit} = 2$, respectively, since 70\% of the events can be discriminated between eruptive and confined flares (see the diagram of Figure 11 in~\citet{DuanA2019}). In this study, we also found two parameters, i.e., $|H_J/\phi^{\prime2}|$ and $E_F/E_P$, with their thresholds are $|H_J/\phi^{\prime2}|_\texttt{crit} = 0.009$ and $(E_F/E_P)_\texttt{crit} = 0.27$, as all the events above $|H_J/\phi^{\prime2}|_\texttt{crit}$ and 95\% of the events above $(E_F/E_P)_\texttt{crit}$ erupted. Furthermore, by such criterion, over 75\% of the events can be discriminated between eruptive and confined flares, which is slightly more successful than the previous study. By further considering that the calculation of the global parameters of the coronal field are much easier than inspecting the detailed magnetic configurations (e.g., searching the MFRs, identifying their axis, and computing the twist number and the  decay index, which needs a significantly larger amount of efforts than this work), the comparison of these two different studies suggests that it might be more effective to predict the eruptive and confined potential of ARs by using the global parameters than studying the specific configuration.

\acknowledgments This work is supported by National Natural Science Foundation of China (NSFC) U2031108 and
Guangdong Basic and Applied Basic Research Foundation (2021A1515011430),
as well as the Fundamental Research Funds for the Central Universities, Sun Yat-sen University (22qntd1912). C.J. acknowledges
support by NSFC 42174200, the Fundamental Research Funds for the Central Universities
(Grant No. HIT.OCEF.2021033), and Shenzhen Technology Project (Grant No. RCJC20210609104422048). The computational work of the NLFFF
extrapolations was carried out on TianHe-1(A), National Supercomputer
Center in Tianjin, China. Data from observations are courtesy of
NASA/SDO.


\begin{thebibliography}{}
\expandafter\ifx\csname natexlab\endcsname\relax\def\natexlab#1{#1}\fi
\providecommand{\url}[1]{\href{#1}{#1}}
\providecommand{\dodoi}[1]{doi:~\href{http://doi.org/#1}{\nolinkurl{#1}}}
\providecommand{\doeprint}[1]{\href{http://ascl.net/#1}{\nolinkurl{http://ascl.net/#1}}}
\providecommand{\doarXiv}[1]{\href{https://arxiv.org/abs/#1}{\nolinkurl{https://arxiv.org/abs/#1}}}

\bibitem[{{Andrews}(2003)}]{Andrews2003}
{Andrews}, M.~D. 2003, \solphys, 218, 261,
  \dodoi{10.1023/B:SOLA.0000013039.69550.bf}

\bibitem[{Aschwanden {et~al.}(2014{\natexlab{a}})Aschwanden, Sun, \&
  Liu}]{Aschwanden2014}
Aschwanden, M.~J., Sun, X., \& Liu, Y. 2014{\natexlab{a}}, The Astrophysical
  Journal, 785, 34.
\newblock \url{http://stacks.iop.org/0004-637X/785/i=1/a=34}

\bibitem[{Aschwanden {et~al.}(2014{\natexlab{b}})Aschwanden, Xu, \&
  Jing}]{Aschwanden2014a}
Aschwanden, M.~J., Xu, Y., \& Jing, J. 2014{\natexlab{b}}, The Astrophysical
  Journal, 797, 50, \dodoi{10.1088/0004-637x/797/1/50}

\bibitem[{{Baumgartner} {et~al.}(2018){Baumgartner}, {Thalmann}, \&
  {Veronig}}]{Baumgartner2018}
{Baumgartner}, C., {Thalmann}, J.~K., \& {Veronig}, A.~M. 2018, in EGU General
  Assembly Conference Abstracts, EGU General Assembly Conference Abstracts,
  5038

\bibitem[{{Benz}(2017)}]{Benz2017}
{Benz}, A.~O. 2017, Living Reviews in Solar Physics, 14, 2,
  \dodoi{10.1007/s41116-016-0004-3}

\bibitem[{Berger(1999)}]{Berger1999}
Berger, M.~A. 1999, Plasma Physics and Controlled Fusion, 41, B167,
  \dodoi{10.1088/0741-3335/41/12b/312}

\bibitem[{{Berger} \& {Field}(1984)}]{Berger1984}
{Berger}, M.~A., \& {Field}, G.~B. 1984, Journal of Fluid Mechanics, 147, 133,
  \dodoi{10.1017/S0022112084002019}

\bibitem[{{Bleybel} {et~al.}(2002){Bleybel}, {Amari}, {van Driel-Gesztelyi}, \&
  {Leka}}]{Bleybel2002}
{Bleybel}, A., {Amari}, T., {van Driel-Gesztelyi}, L., \& {Leka}, K.~D. 2002,
  \aap, 395, 685, \dodoi{10.1051/0004-6361:20021332}

\bibitem[{{Bobra} {et~al.}(2014){Bobra}, {Sun}, {Hoeksema}, {Turmon}, {Liu},
  {Hayashi}, {Barnes}, \& {Leka}}]{Bobra2014}
{Bobra}, M.~G., {Sun}, X., {Hoeksema}, J.~T., {et~al.} 2014, \solphys, 289,
  3549, \dodoi{10.1007/s11207-014-0529-3}

\bibitem[{{DeRosa} {et~al.}(2009){DeRosa}, {Schrijver}, {Barnes}, {Leka},
  {Lites}, {Aschwanden}, {Amari}, {Canou}, {McTiernan}, {R{\'e}gnier},
  {Thalmann}, {Valori}, {Wheatland}, {Wiegelmann}, {Cheung}, {Conlon},
  {Fuhrmann}, {Inhester}, \& {Tadesse}}]{DeRosa2009}
{DeRosa}, M.~L., {Schrijver}, C.~J., {Barnes}, G., {et~al.} 2009, \apj, 696,
  1780, \dodoi{10.1088/0004-637X/696/2/1780}

\bibitem[{Duan {et~al.}(2019)Duan, Jiang, He, Feng, Zou, \& Cui}]{DuanA2019}
Duan, A., Jiang, C., He, W., {et~al.} 2019, The Astrophysical Journal, 884, 73,
  \dodoi{10.3847/1538-4357/ab3e33}

\bibitem[{{Duan} {et~al.}(2017){Duan}, {Jiang}, {Hu}, {Zhang}, {Gary}, {Wu}, \&
  {Cao}}]{DuanA2017}
{Duan}, A., {Jiang}, C., {Hu}, Q., {et~al.} 2017, \apj, 842, 119,
  \dodoi{10.3847/1538-4357/aa76e1}

\bibitem[{{Duan} {et~al.}(2021){Duan}, {Jiang}, {Zhou}, {Feng}, \&
  {Cui}}]{DuanA2021apjl}
{Duan}, A., {Jiang}, C., {Zhou}, Z., {Feng}, X., \& {Cui}, J. 2021, \apjl, 907,
  L23, \dodoi{10.3847/2041-8213/abd638}

\bibitem[{Emslie {et~al.}(2012)Emslie, Dennis, Shih, Chamberlin, Mewaldt,
  Moore, Share, Vourlidas, \& Welsch}]{Emslie2012}
Emslie, A.~G., Dennis, B.~R., Shih, A.~Y., {et~al.} 2012, The Astrophysical
  Journal, 759, 71, \dodoi{10.1088/0004-637X/759/1/71}

\bibitem[{{Finn}(1984)}]{Finn1984}
{Finn}, J.~M. 1984, Comments Plasma Phys. Controlled Fusion, 9, 111

\bibitem[{{Fletcher} {et~al.}(2011){Fletcher}, {Dennis}, {Hudson}, {Krucker},
  {Phillips}, {Veronig}, {Battaglia}, {Bone}, {Caspi}, {Chen}, {Gallagher},
  {Grigis}, {Ji}, {Liu}, {Milligan}, \& {Temmer}}]{Fletcher2011}
{Fletcher}, L., {Dennis}, B.~R., {Hudson}, H.~S., {et~al.} 2011, \ssr, 159, 19,
  \dodoi{10.1007/s11214-010-9701-8}

\bibitem[{Gilchrist {et~al.}(2012)Gilchrist, Wheatland, \&
  Leka}]{Gilchrist2012}
Gilchrist, S.~A., Wheatland, M.~S., \& Leka, K.~D. 2012, Solar Physics, 276,
  133, \dodoi{10.1007/s11207-011-9878-3}

\bibitem[{{Gupta} {et~al.}(2021){Gupta}, {Thalmann}, \& {Veronig}}]{Gupta2021}
{Gupta}, M., {Thalmann}, J.~K., \& {Veronig}, A.~M. 2021, \aap, 653, A69,
  \dodoi{10.1051/0004-6361/202140591}

\bibitem[{{Hagino} \& {Sakurai}(2004)}]{Hagino2004}
{Hagino}, M., \& {Sakurai}, T. 2004, \pasj, 56, 831,
  \dodoi{10.1093/pasj/56.5.831}

\bibitem[{{Hoeksema} {et~al.}(2014){Hoeksema}, {Liu}, {Hayashi}, {Sun},
  {Schou}, {Couvidat}, {Norton}, {Bobra}, {Centeno}, {Leka}, {Barnes}, \&
  {Turmon}}]{Hoeksema2014}
{Hoeksema}, J.~T., {Liu}, Y., {Hayashi}, K., {et~al.} 2014, \solphys, 289,
  3483, \dodoi{10.1007/s11207-014-0516-8}

\bibitem[{{James} {et~al.}(2018){James}, {Valori}, {Green}, {Liu}, {Cheung},
  {Guo}, \& {van Driel-Gesztelyi}}]{James2018}
{James}, A.~W., {Valori}, G., {Green}, L.~M., {et~al.} 2018, \apjl, 855, L16,
  \dodoi{10.3847/2041-8213/aab15d}

\bibitem[{{Jiang} \& {Feng}(2013)}]{JiangC2013NLFFF}
{Jiang}, C., \& {Feng}, X. 2013, \apj, 769, 144,
  \dodoi{10.1088/0004-637X/769/2/144}

\bibitem[{{Jiang} \& {Feng}(2014)}]{Jiang2014Prep}
---. 2014, \solphys, 289, 63, \dodoi{10.1007/s11207-013-0346-0}

\bibitem[{Jing {et~al.}(2009)Jing, Chen, Wiegelmann, Xu, Park, \&
  Wang}]{Jingj2009}
Jing, J., Chen, P.~F., Wiegelmann, T., {et~al.} 2009, The Astrophysical
  Journal, 696, 84, \dodoi{10.1088/0004-637x/696/1/84}

\bibitem[{Jing {et~al.}(2018)Jing, Liu, Lee, Ji, Liu, Xu, \& Wang}]{JingJ2018}
Jing, J., Liu, C., Lee, J., {et~al.} 2018, The Astrophysical Journal, 864, 138,
  \dodoi{10.3847/1538-4357/aad6e4}

\bibitem[{{Jing} {et~al.}(2010){Jing}, {Tan}, {Yuan}, {Wang}, {Wiegelmann},
  {Xu}, \& {Wang}}]{JingJ2010}
{Jing}, J., {Tan}, C., {Yuan}, Y., {et~al.} 2010, \apj, 713, 440,
  \dodoi{10.1088/0004-637X/713/1/440}

\bibitem[{{Kazachenko} {et~al.}(2017){Kazachenko}, {Lynch}, {Welsch}, \&
  {Sun}}]{Kazachenko2017}
{Kazachenko}, M.~D., {Lynch}, B.~J., {Welsch}, B.~T., \& {Sun}, X. 2017, \apj,
  845, 49, \dodoi{10.3847/1538-4357/aa7ed6}

\bibitem[{LaBonte {et~al.}(2007)LaBonte, Georgoulis, \& Rust}]{LaBonte2007}
LaBonte, B.~J., Georgoulis, M.~K., \& Rust, D.~M. 2007, The Astrophysical
  Journal, 671, 955, \dodoi{10.1086/522682}

\bibitem[{{Leka} {et~al.}(1993){Leka}, {Canfield}, {McClymont}, {de La
  Beaujardiere}, {Fan}, \& {Tang}}]{Leka1993}
{Leka}, K.~D., {Canfield}, R.~C., {McClymont}, A.~N., {et~al.} 1993, \apj, 411,
  370, \dodoi{10.1086/172837}

\bibitem[{{Li} {et~al.}(2021){Li}, {Chen}, {Hou}, {Veronig}, {Yang}, \&
  {Zhang}}]{LiT2021}
{Li}, T., {Chen}, A., {Hou}, Y., {et~al.} 2021, \apjl, 917, L29,
  \dodoi{10.3847/2041-8213/ac1a15}

\bibitem[{{Li} {et~al.}(2020){Li}, {Hou}, {Yang}, {Zhang}, {Liu}, \&
  {Veronig}}]{LiT2020}
{Li}, T., {Hou}, Y., {Yang}, S., {et~al.} 2020, \apj, 900, 128,
  \dodoi{10.3847/1538-4357/aba6ef}

\bibitem[{{Linan} {et~al.}(2018){Linan}, {Pariat}, {Moraitis}, {Valori}, \&
  {Leake}}]{Linan2018}
{Linan}, L., {Pariat}, {\'E}., {Moraitis}, K., {Valori}, G., \& {Leake}, J.
  2018, \apj, 865, 52, \dodoi{10.3847/1538-4357/aadae7}

\bibitem[{{Liu} {et~al.}(2023){Liu}, {Welsch}, {Valori}, {Georgoulis}, {Guo},
  {Pariat}, {Park}, \& {Thalmann}}]{LiuY2023}
{Liu}, Y., {Welsch}, B.~T., {Valori}, G., {et~al.} 2023, \apj, 942, 27,
  \dodoi{10.3847/1538-4357/aca3a6}

\bibitem[{{Metcalf} {et~al.}(2005){Metcalf}, {Leka}, \& {Mickey}}]{Metcalf2005}
{Metcalf}, T.~R., {Leka}, K.~D., \& {Mickey}, D.~L. 2005, \apjl, 623, L53,
  \dodoi{10.1086/429961}

\bibitem[{{Moraitis} {et~al.}(2019){Moraitis}, {Sun}, {Pariat}, \&
  {Linan}}]{Moraitis2019B}
{Moraitis}, K., {Sun}, X., {Pariat}, {\'E}., \& {Linan}, L. 2019, \aap, 628,
  A50, \dodoi{10.1051/0004-6361/201935870}

\bibitem[{Nindos \& Andrews(2004)}]{Nindos2004}
Nindos, A., \& Andrews, M.~D. 2004, The Astrophysical Journal, 616, L175,
  \dodoi{10.1086/426861}

\bibitem[{{Pariat} {et~al.}(2017){Pariat}, {Leake}, {Valori}, {Linton},
  {Zuccarello}, \& {Dalmasse}}]{Pariat2017}
{Pariat}, E., {Leake}, J.~E., {Valori}, G., {et~al.} 2017, \aap, 601, A125,
  \dodoi{10.1051/0004-6361/201630043}

\bibitem[{Pesnell {et~al.}(2012)Pesnell, Thompson, \& Chamberlin}]{Pesnell2012}
Pesnell, W.~D., Thompson, B.~J., \& Chamberlin, P.~C. 2012, Solar Physics, 275,
  3, \dodoi{10.1007/s11207-011-9841-3}

\bibitem[{{Pevtsov} {et~al.}(1994){Pevtsov}, {Canfield}, \&
  {Metcalf}}]{Pevtsov1994}
{Pevtsov}, A.~A., {Canfield}, R.~C., \& {Metcalf}, T.~R. 1994, \apjl, 425,
  L117, \dodoi{10.1086/187324}

\bibitem[{R{\'e}gnier(2013)}]{Regnier2013}
R{\'e}gnier, S. 2013, Solar Physics, 288, 481,
  \dodoi{10.1007/s11207-013-0367-8}

\bibitem[{Smyrli {et~al.}(2010)Smyrli, Zuccarello, Romano, Zuccarello,
  Guglielmino, Spadaro, Hood, \& Mackay}]{Smyrli2010}
Smyrli, A., Zuccarello, F., Romano, P., {et~al.} 2010, Astronomy and
  Astrophysics, 521, A56, \dodoi{10.1051/0004-6361/200913275}

\bibitem[{Sun {et~al.}(2015)Sun, Bobra, Hoeksema, Liu, Li, Shen, Couvidat,
  Norton, \& Fisher}]{SunXD2015}
Sun, X., Bobra, M.~G., Hoeksema, J.~T., {et~al.} 2015, \apjl, 804, L28

\bibitem[{Thalmann {et~al.}(2019{\natexlab{a}})Thalmann, Linan, Pariat, \&
  Valori}]{Thalmann2019A}
Thalmann, J.~K., Linan, L., Pariat, E., \& Valori, G. 2019{\natexlab{a}}, The
  Astrophysical Journal, 880, L6, \dodoi{10.3847/2041-8213/ab2e73}

\bibitem[{Thalmann {et~al.}(2019{\natexlab{b}})Thalmann, Moraitis, Linan,
  Pariat, Valori, \& Dalmasse}]{Thalmann2019B}
Thalmann, J.~K., Moraitis, K., Linan, L., {et~al.} 2019{\natexlab{b}}, The
  Astrophysical Journal, 887, 64, \dodoi{10.3847/1538-4357/ab4e15}

\bibitem[{{Thalmann} {et~al.}(2020){Thalmann}, {Sun}, {Moraitis}, \&
  {Gupta}}]{Thalmann2020}
{Thalmann}, J.~K., {Sun}, X., {Moraitis}, K., \& {Gupta}, M. 2020, \aap, 643,
  A153, \dodoi{10.1051/0004-6361/202038921}

\bibitem[{Tian {et~al.}(2002)Tian, Wang, \& Wu}]{TianLR2002}
Tian, L., Wang, J., \& Wu, D. 2002, Solar Physics, 209, 375,
  \dodoi{10.1023/A:1021201817701}

\bibitem[{{Toriumi} {et~al.}(2017){Toriumi}, {Schrijver}, {Harra}, {Hudson}, \&
  {Nagashima}}]{Toriumi2017}
{Toriumi}, S., {Schrijver}, C.~J., {Harra}, L.~K., {Hudson}, H., \&
  {Nagashima}, K. 2017, \apj, 834, 56, \dodoi{10.3847/1538-4357/834/1/56}

\bibitem[{Tziotziou {et~al.}(2012)Tziotziou, Georgoulis, \&
  Raouafi}]{Tziotziou2012}
Tziotziou, K., Georgoulis, M.~K., \& Raouafi, N.-E. 2012, The Astrophysical
  Journal, 759, L4, \dodoi{10.1088/2041-8205/759/1/L4}

\bibitem[{Valori {et~al.}(2012)Valori, D{\'e}moulin, \& Pariat}]{Valori2012B}
Valori, G., D{\'e}moulin, P., \& Pariat, E. 2012, Solar Physics, 278, 347,
  \dodoi{10.1007/s11207-012-9951-6}

\bibitem[{Valori {et~al.}(2013)Valori, D{\'e}moulin, Pariat, \&
  Masson}]{Valori2013}
Valori, G., D{\'e}moulin, P., Pariat, E., \& Masson, S. 2013, Astronomy \&
  Astrophysics, 553, A38, \dodoi{10.1051/0004-6361/201220982}

\bibitem[{Vasantharaju {et~al.}(2018)Vasantharaju, Vemareddy, Ravindra, \&
  Doddamani}]{Vasantharaju2018}
Vasantharaju, N., Vemareddy, P., Ravindra, B., \& Doddamani, V.~H. 2018, The
  Astrophysical Journal, 860, 58, \dodoi{10.3847/1538-4357/aac272}

\bibitem[{{Wang} {et~al.}(1994){Wang}, {Ewell}, {Zirin}, \&
  {Ai}}]{WangHM1994ApJ}
{Wang}, H., {Ewell}, M.~W., J., {Zirin}, H., \& {Ai}, G. 1994, \apj, 424, 436,
  \dodoi{10.1086/173901}

\bibitem[{Wang {et~al.}(1994)Wang, Xu, \& Zhang}]{WangHM1994SP}
Wang, T., Xu, A., \& Zhang, H. 1994, Solar Physics, 155, 99,
  \dodoi{10.1007/BF00670733}

\bibitem[{{Wang} \& {Zhang}(2007)}]{WangYM2007}
{Wang}, Y., \& {Zhang}, J. 2007, in American Astronomical Society Meeting
  Abstracts, Vol. 210, American Astronomical Society Meeting Abstracts \#210,
  29.17

\bibitem[{{Wiegelmann}(2004)}]{Wiegelmann2004}
{Wiegelmann}, T. 2004, \solphys, 219, 87,
  \dodoi{10.1023/B:SOLA.0000021799.39465.36}

\bibitem[{{Wiegelmann} {et~al.}(2017){Wiegelmann}, {Petrie}, \&
  {Riley}}]{Wiegelmann2017}
{Wiegelmann}, T., {Petrie}, G. J.~D., \& {Riley}, P. 2017, \ssr, 210, 249,
  \dodoi{10.1007/s11214-015-0178-3}

\bibitem[{{Yashiro} {et~al.}(2006){Yashiro}, {Akiyama}, {Gopalswamy}, \&
  {Howard}}]{Yashiro2006}
{Yashiro}, S., {Akiyama}, S., {Gopalswamy}, N., \& {Howard}, R.~A. 2006, \apjl,
  650, L143, \dodoi{10.1086/508876}

\bibitem[{Zuccarello {et~al.}(2018)Zuccarello, Pariat, Valori, \&
  Linan}]{Zuccarello2018}
Zuccarello, F.~P., Pariat, E., Valori, G., \& Linan, L. 2018, The Astrophysical
  Journal, 863, 41, \dodoi{10.3847/1538-4357/aacdfc}

\bibitem[{Zuccarello {et~al.}(2011)Zuccarello, Romano, Zuccarello, \&
  Poedts}]{Zuccarello2011}
Zuccarello, F.~P., Romano, P., Zuccarello, F., \& Poedts, S. 2011, Astronomy \&
  Astrophysics, 530, A36, \dodoi{10.1051/0004-6361/201116700}

\end{thebibliography}
\end{document}